\documentclass[preprint,pre]{revtex4}
\usepackage{bm}
\usepackage{amsmath}
\usepackage{relsize}
\renewcommand{\vec}[1]{\bm{#1}}

\usepackage{graphicx}
\begin{document}
\title{A Thermostat for Molecular Dynamics of Complex Fluids}
\author{Michael P. Allen}
\thanks{Invited Talk at Symposium on Progress and Future Prospects in Molecular
  Dynamics Simulation -- In Memory of Professor Shuichi Nos\'{e}}
\affiliation{Department of Physics and Centre for Scientific Computing \\
University of Warwick, Coventry CV4 7AL, United Kingdom}
\author{Friederike Schmid}
\affiliation{Fakult\"{a}t f\"{u}r Physik, Universit\"{a}t Bielefeld,
D-33615 Bielefeld, Germany}
\begin{abstract}
  A thermostat of the Nos\'{e}-Hoover type, based on relative velocities and a
  local definition of the temperature, is presented. The thermostat is
  momentum-conserving and Galilean-invariant, which should make it suitable for
  use in Dissipative Particle Dynamics simulations, as well as nonequilibrium
  molecular dynamics simulations.
\end{abstract}
\maketitle
\section{Introduction}
\label{sec:intro}
The original papers of \citet{nose.s:1984.a,nose.s:1984.b} provided a new
perspective on the generation of statistical ensembles by dynamical simulation.
They showed that a deterministic set of equations of motion, involving just one
or two extra degrees of freedom, can sample configurations from the canonical
ensemble. This complements the stochastic method of \citet{andersen.hc:1980.a},
which generates the canonical ensemble by periodic reselection of velocities
from the Maxwell-Boltzmann distribution.  The work of
\citet{hoover.wg:1985.a,hoover.wg:1991.a} further clarified the nature of the
isothermal dynamical equations and how to derive them (see also the contribution
of Hoover to these proceedings). The Nos\'{e}-Hoover equations incorporate a
dynamical friction coefficient, whose fluctuations are driven by the difference
between the instantaneous kinetic temperature (defined through the sum of
squares of particle velocities) and the desired temperature.

This paper presents a new thermostat of the Nos\'{e}-Hoover type, based on an
instantaneous temperature which is calculated as a weighted sum of squares of
relative velocities of atom pairs. The frictional term in the equations of
motion also enters in pairwise fashion, conserving momentum, and making the
dynamics invariant to a Galilean transformation of velocities. There are two
areas in which such a thermostat may be useful. The first area is nonequilibrium
molecular dynamics (NEMD). With a conventional Nos\'{e}-Hoover thermostat, it is
necessary to apply the friction term to the \emph{peculiar} velocities of the
particles, i.e.\ the difference between the true velocities and the local
streaming velocity of the fluid at the particle positions. Failure to do this
can lead to unphysical (thermostat-induced) behaviour
\cite{delhommelle.j:2003.a} such as the stabilisation of string phases
\cite{evans.dj:1986.a,evans.dj:1992.a} or the generation of steady-state
antisymmetric stress \cite{travis.kp:1995.b}. One solution to these problems is
to use a profile-unbiased thermostat, which requires a self-consistent
determination of the streaming velocity in the course of the simulation
\cite{evans.dj:1986.a,evans.dj:1992.a,travis.kp:1995.b}. A thermostat based on
the relative velocities of nearby pairs of atoms may avoid, or at least
ameliorate, the problem. The second area of application of the pairwise
thermostat is dissipative particle dynamics (DPD). Here, in order to preserve
hydrodynamic behaviour, it is essential for any thermostat to conserve momentum,
and a pairwise form is one way of achieving this. This paper concentrates on the
DPD case, since the suggestion of using a pairwise Nos\'{e}-Hoover thermostat
was first made in this context by \citet{stoyanov.sd:2005.a}. However, it should
be borne in mind that the thermostat may be applied equally well to, e.g.\ 
Lennard-Jones fluids.

The paper is organised as follows. Section \ref{sec:dpd} contains a brief
summary of DPD, concentrating on the temperature control aspects. Section
\ref{sec:thermostat} derives the equations of motion for the pairwise
Nos\'{e}-Hoover thermostat, and also summarizes the equations for a thermostat
based on the configurational temperature, due to \citet{braga.c:2005.a}, for
comparison. Section \ref{sec:results} presents the results of some preliminary
tests for DPD simulations. Finally, section \ref{sec:discussion} contains the
conclusions.
\section{Dissipative Particle Dynamics}
\label{sec:dpd}
DPD \cite{hoogerbrugge.pj:1992.a,koelman.jmva:1993.a} has become a popular tool
for simulating the behaviour of both simple and complex fluids. It consists of
the solution of the classical equations of motion for a system of interacting
particles, together with a set of stochastic and dissipative forces which
control the temperature and allow one to choose the viscosity. For a simple
fluid the equations may be written
\cite{hoogerbrugge.pj:1992.a,koelman.jmva:1993.a,espanol.p:1995.b}
\begin{subequations}
\label{eqn:dpd}
\begin{align}
\dot{\vec{r}}_i &= \vec{v}_i = \vec{p}_i/m_i
\label{eqn:dpd.a}
\\
\dot{\vec{p}}_i &= \vec{f}_{i}(\vec{r}) -\xi \vec{V}_{i}(\vec{r},\vec{p}) +
\sigma\vec{R}_{i}(\vec{r},\vec{p}) \;,
\label{eqn:dpd.b}
\end{align}
\end{subequations}
where $\vec{r}$ and $\vec{p}$ stand for the complete set of coordinates and
momenta.  The so-called conservative forces $\vec{f}_{i}$ are derived from a
pair-potential term in the Hamiltonian $\vec{f}_{i} = -(\partial H/\partial
\vec{r}_i)$ and so may be written as $\vec{f}_{i} = \sum_{j\neq i}
\vec{f}_{ij}$, with $\vec{f}_{ji}=-\vec{f}_{ij}$.  In DPD these pair forces
usually take the form
\begin{subequations}
\label{eqn:conforce}
\begin{align}
 \vec{f}_{ij} &= \alpha \vec{w}_{ij}  = \alpha \vec{w}(\vec{r}_{ij}) \;,
\qquad\text{with}\qquad
\vec{w}(\vec{r}) = w(r) \hat{\vec{r}} \\
\text{and}\quad
w(r) &= 
\begin{cases} 
\bigl(1-r/r_\text{c}\bigr) & r\leq r_\text{c} \\ 
0 & r> r_\text{c} 
\end{cases} \;.
\end{align}
\end{subequations}
Here $\vec{r}_{ij}=\vec{r}_{i}-\vec{r}_{j}$, $r = |\vec{r}|$, $\hat{\vec{r}} =
\vec{r}/r$. The parameter $\alpha$ determines the strength of the conservative
interactions, and $r_\text{c}$ is the cutoff.  

The dissipative forces $-\xi\vec{V}_{i}$ are also written in pairwise fashion
$\vec{V}_{i} =\sum_{j\neq i}\vec{V}_{ij}$ with $\vec{V}_{ji}=-\vec{V}_{ij}$,
usually defined thus:
\begin{equation}
\label{eqn:F}
\vec{V}_{ij} = \bigl(\vec{v}_{ij}\cdot\vec{w}_{ij}\bigr)
\vec{w}_{ij} = w(r_{ij})^2 \bigl(\vec{v}_{ij}\cdot\hat{\vec{r}}_{ij}\bigr)
\hat{\vec{r}}_{ij}
\end{equation}
where $\vec{v}_{ij}=\vec{v}_{i}-\vec{v}_{j}$.  A choice has been made here to
use the same weighting function $\vec{w}(\vec{r})$ as in the specification of
conservative forces.  $\sigma\vec{R}_{i}$ is short for the random ``forces'',
which also act between pairs, with a weight function $\vec{w}(\vec{r})$; the
strength parameter $\sigma$ is related through the fluctuation-dissipation
theorem to the friction coefficient $\xi$ and the temperature $k_\text{B}T$ (see
\cite{hoogerbrugge.pj:1992.a,koelman.jmva:1993.a,espanol.p:1995.b} for more
details).  The pairwise nature of all these forces guarantees the momentum
conservation necessary to ensure hydrodynamic behaviour: in other words, the
dynamics is Galilean-invariant.  The particles represent fluid regions, rather
than individual atoms and molecules: the softness and simplicity of the
interactions permit the use of a long time step, compared with conventional
molecular dynamics. This, and the acceleration of physical processes compared
with those seen in more realistic simulations, gives an advantage of several
orders of magnitude, at the cost of a very rough mapping onto specific molecular
properties.

A slightly more general view of DPD treats it as conventional molecular dynamics
using soft potentials, supplemented by a momentum-conserving thermostat which
acts between pairs. 
\citet{lowe.cp:1999.a} takes this approach, rather than solving the above
equations. Instead, each timestep $\Delta t$ involves the following operations.
\begin{enumerate}
\item Positions and momenta are advanced using 
$\dot{\vec{r}}_i = \vec{p}_i/m_i$, 
$\dot{\vec{p}}_i = \vec{f}_{i}$.
\item Every pair $ij$ (in random order, and possibly subject to a distance
  dependent weight or range function) is examined and, with probability
  $P=\nu\Delta t$, the momenta are updated: $\vec{p}_i := \vec{p}_i +
  \Delta\vec{p}_{ij}$, $\vec{p}_j := \vec{p}_j - \Delta\vec{p}_{ij}$, with
\begin{equation*}
\Delta\vec{p}_{ij} = m_{ij}\left[
\zeta \sqrt{k_\text{B}T/m_{ij}}
- (\vec{v}_{ij}\cdot\hat{\vec{r}}_{ij}) \right]\hat{\vec{r}}_{ij}
\end{equation*}
where $\zeta$ is picked from a Gaussian distribution with zero mean and unit
variance, and $m_{ij} = m_im_j/(m_i+m_j)$.
\end{enumerate}
This procedure periodically reselects the component of the relative velocity
along $\hat{\vec{r}}_{ij}$ from the Maxwell-Boltzmann distribution corresponding
to reduced mass $m_{ij}$.  The key parameter is the stochastic randomization
frequency $\nu$: high values of $\nu$ give effective temperature control, but
also a high viscosity; low values give very weak temperature control while
allowing the viscosity to be low. The thermostat is closely related to the one
originally proposed by \citet{andersen.hc:1980.a}.

Recently, \citet{stoyanov.sd:2005.a} have proposed a modification of the above
method: the fraction $(1-P)$ of pairs which do not have their relative
velocities stochastically updated, are instead thermalized by a deterministic
method. For each such pair, a dissipative force is calculated and used to
correct the momenta during the deterministic part of the step, incorporating a
temperature-dependent controlling factor. Finally, the Lowe velocity reselection
process is applied to the remaining fraction $P$ of pairs as before.  The idea
of Stoyanov and Groot is to give more control over the separate effects of
thermalization, namely temperature control and changing viscosity.
\citet{stoyanov.sd:2005.a} call the deterministic part of their thermostat
``Nos\'{e}-Hoover'', but actually it is not of this form, and has not been shown
to generate the canonical ensemble. It may be noted that an algorithm
resembling that of Nos\'{e} and Hoover was also described by
\citet{besold.g:2000.a}.
\section{Pairwise Nos\'{e}-Hoover Thermostat}
\label{sec:thermostat}
\subsection{Derivation of Equations of Motion}
\label{sec:derivation}
The purpose of this paper is to present a Galilean-invariant thermostat of the
Nos\'{e}-Hoover type, which generates the canonical ensemble. The derivation is
a straightforward implementation of the approach of \citet{hoover.wg:1991.a},
and a special case of the generalized Nos\'{e}-Hoover equations discussed by
\citet{kusnezov.d:1990.a} and \citet{martyna.gj:1992.a}.  The result is assumed
to be of the form
\begin{subequations}
\label{eqn:nh}
\begin{align}
\dot{\vec{r}}_i &= \vec{p}_i/m_i
\label{eqn:nh.a}
\\
\dot{\vec{p}}_i
&=  \vec{f}_{i}(\vec{r}) -\xi \vec{V}_i(\vec{r},\vec{p})
\label{eqn:nh.b}
\\
\dot{\xi} &= G_\xi(\vec{r},\vec{p})
\label{eqn:nh.c}
\end{align}
\end{subequations}
with the $\vec{V}_i(\vec{r},\vec{p})$ given by eqn~\eqref{eqn:F}.
Eqns~\eqref{eqn:nh.a} and \eqref{eqn:nh.b} are written down by analogy with
eqns~\eqref{eqn:dpd}.  The random forces are dropped, the friction coefficient
$\xi$ is now an additional \emph{dynamical} variable, and the right-hand side
of eqn~\eqref{eqn:nh.c} is the object of the derivation.  This is obtained from
the generalized Liouville equation for the (stationary) phase space distribution
function $\varrho(\vec{r},\vec{p},\xi)$
\begin{equation}
\label{eqn:liouville}
\sum_i \frac{\partial}{\partial\vec{r}_i} \cdot \bigl(\rho\dot{\vec{r}}_i\bigr)
+
\sum_i \frac{\partial}{\partial\vec{p}_i} \cdot \bigl(\rho\dot{\vec{p}}_i\bigr)
+
 \frac{\partial}{\partial\xi} \bigl(\rho\dot{\xi}\bigr)
= 0 \;.
\end{equation}
The ansatz is made that $G_\xi(\vec{r},\vec{p})$ in eqn~\eqref{eqn:nh.c} depends
only on positions and momenta, so $\partial\dot{\xi}/\partial\xi=0$.  Direct
substitution shows that equation \eqref{eqn:liouville} is satisfied by the
product form
\begin{equation*}
\rho(\vec{r},\vec{p},\xi) \propto 
\exp\bigl\{-H(\vec{r},\vec{p})/k_\text{B}T\bigr\} 
\exp\bigl\{-\tfrac{1}{2}Q_{\xi}\xi^2/k_\text{B}T\bigr\}
\end{equation*}
where $Q_{\xi}$ is an arbitrary constant, provided
\begin{align}
G_{\xi}(\vec{r},\vec{p})
&= Q_{\xi}^{-1}  \sum_i \left( \frac{\vec{p}_i}{m_i}\cdot\vec{V}_i 
- k_\text{B}T \frac{\partial}{\partial\vec{p}_i}\cdot\vec{V}_i \right)
\notag \\
&= Q_{\xi}^{-1}  \sum_i\sum_{j\neq i}
 \bigl( \vec{v}_{i}\cdot\vec{V}_{ij}
- (k_\text{B}T/m_i) w(r_{ij})^2 \bigr)
\notag \\
&= Q_{\xi}^{-1}  \sum_i\sum_{j<i}
 \bigl( \vec{v}_{ij}\cdot\vec{V}_{ij}
- (k_\text{B}T/m_{ij}) w(r_{ij})^2  \bigr)
\notag \\
&= Q_{\xi}^{-1}  \sum_i\sum_{j<i} w(r_{ij})^2
 \bigl[ \bigl(\vec{v}_{ij}\cdot\hat{\vec{r}}_{ij}\bigr)^2 
- k_\text{B}T/m_{ij}  \bigr] \:.
\label{eqn:G}
\end{align}
Once more, the reduced mass $m_{ij}$ appears.  The term in square brackets
vanishes if an average is taken over the canonical momentum distribution. The
equation has a straightforward physical interpretation, acting to damp the
difference between the instantaneous temperature corresponding to the component
of relative velocity $\vec{v}_{ij}$ along the inter-particle vector, and the
canonical ensemble average of this quantity. The prefactor $Q_{\xi}$ controls the
``thermal inertia'' in the same way as the corresponding parameter in the
conventional Nos\'{e}-Hoover method, and the function $w$ gives a
higher weighting to closer pairs. There is a conserved ``energy function''
\begin{equation}
H_\xi(\vec{r},\vec{p},\xi,\varphi_\xi) = H(\vec{r},\vec{p}) + \tfrac{1}{2}Q_{\xi} \xi^2 + \varphi_\xi 
\qquad\text{where}\quad \dot{\varphi}_\xi = \xi k_\text{B}T \sum_{i<j}
w(r_{ij})^2/m_{ij} 
\label{eqn:engxi}
\end{equation}
as may be checked by time differentiation and direct substitution of the
equations of motion.
\subsection{Integration Algorithm}
\label{sec:algorithm}
It is not the aim here to discuss the optimal algorithm for integration of the
equations of motion \eqref{eqn:nh}. Instead, the simplest modified
velocity-Verlet algorithm \citep{martyna.gj:1994.a}, that is commonly used in
DPD \cite{besold.g:2000.a}, is adopted:
\begin{subequations}
\begin{align}
\tilde{\vec{p}}_i := \vec{p}_i & := \vec{p}_i + \tfrac{1}{2}\Delta t \,
\left(\vec{f}_i - \xi \vec{V}_i \right)
&& \vec{p}_i(\tfrac{1}{2}\Delta t)
\\
\tilde{\xi} := \xi & := \xi + \tfrac{1}{2}\Delta t\, G_\xi
&& \xi(\tfrac{1}{2}\Delta t)
\\
\vec{r}_i & := \vec{r}_i + \Delta t \, \vec{p}_i / m_i 
&& \vec{r}_i(\Delta t)
\\
\vec{f}_i & := \vec{f}_i(\vec{r})
&& \vec{f}_i(\Delta t)
\\
\vec{V}_i & := \vec{V}_i(\vec{r},\vec{p})
&& \vec{V}_i(\Delta t)
\label{alg:e}
\\
G_\xi & := G_\xi(\vec{r},\vec{p})
&& G_\xi(\Delta t)
\label{alg:f}
\\
\vec{p}_i & := \tilde{\vec{p}}_i
+ \tfrac{1}{2}\Delta t \,
\left(\vec{f}_i-\xi\vec{V}_i\right)
&& \vec{p}_i(\Delta t)
\label{alg:g}
\\
\xi &:= \tilde{\xi} + \tfrac{1}{2}\Delta t \, G_\xi
&& \xi(\Delta t)
\label{alg:h}
\end{align}
\end{subequations}
Steps \eqref{alg:e}--\eqref{alg:h} may be iterated to convergence, because the
momenta at time $t+\Delta t$ should be used in the evaluation of $G_\xi$ and
$\vec{V}_i$. However, because of the expense of calculating the pairwise terms,
in DPD it is usual to stop after one evaluation of the expressions above, and
this is the approach adopted here.  Some might prefer a strictly reversible
integrator \citep{martyna.gj:1996.a}, while others favour the Runge-Kutta
method: consideration of these possibilities is deferred.
\subsection{Configurational Nos\'{e}-Hoover Thermostat}
\label{sec:configtherm}
The canonical ensemble result
\begin{equation}
 \sum_j%
\left\langle \left| \frac{\partial U}{\partial\vec{r}_j} \right|^2\right\rangle
=
k_\text{B}T \sum_j\left\langle
\frac{\partial}{\partial\vec{r}_j}\cdot\frac{\partial U}{\partial\vec{r}_j}
\right\rangle
 \;.
\label{eqn:conftemp}
\end{equation}
has been known for many years \citep{hirschfelder.jo:1960.a} and has recently
been used to define a configurational temperature $T_\text{c}$ in simulation
\citep{rugh.hh:1997.a,butler.bd:1998.a} and experiment
\citep{han.yl:2004.a,han.yl:2005.a}. Recently, one of us \citep{allen.mp:2006.a}
has suggested monitoring this quantity as an indicator of lack of equilibrium
due to excessive timesteps in DPD. It is natural to consider applying a
thermostat to control this variable \citep{delhommelle.j:2001.a,braga.c:2005.a}
and here the equations of motion of \citet{braga.c:2005.a} are used:
\begin{subequations}
\label{eqn:nhc}
\begin{align}
\dot{\vec{r}}_i &= \vec{p}_i/m_i + \mu \vec{f}_i(\vec{r})
\label{eqn:nhc.a}
\\
\dot{\vec{p}}_i &=  \vec{f}_{i}(\vec{r})
\label{eqn:nhc.b}
\\
\dot{\mu} &= G_\mu(\vec{r})
\label{eqn:nhc.c}
\end{align}
\end{subequations}
where
\begin{equation}
G_\mu = Q_\mu^{-1}  
\sum_j%
\left(
\left| \frac{\partial U}{\partial\vec{r}_j} \right|^2
-  k_\text{B}T 
\frac{\partial}{\partial\vec{r}_j}\cdot\frac{\partial U}{\partial\vec{r}_j}
\right)
 \:.
\end{equation}
The quantity $\mu$ plays the role of a fluctuating mobility: that is, a
proportionality between force and drift velocity, as seen in the ``position
Langevin equation'' or Schmoluchowski equation \citep{braga.c:2005.a}.  Once
again there is a conserved ``energy function''
\begin{equation}
H_\mu(\vec{r},\vec{p},\mu,\varphi_\mu) 
= H(\vec{r},\vec{p}) + \tfrac{1}{2}Q_{\mu} \mu^2 + \varphi_\mu
\qquad\text{where}\quad \dot{\varphi}_\mu = \mu k_\text{B}T 
\sum_{j} \frac{\partial}{\partial\vec{r}_j}\cdot\frac{\partial
  U}{\partial\vec{r}_j} \:.
\label{eqn:engmu}
\end{equation}
\citet{braga.c:2005.a} have presented a simple integration algorithm for these
equations, which we use here. The canonical distribution may also be shown to be
a steady-state solution of the above equations of motion, and they share with
the thermostatted equations of section \ref{sec:derivation} the property of
Galilean invariance.
%
%
\section{Results}
\label{sec:results}
Tests have been carried out using the standard ``water'' DPD model
\citep{groot.rd:1997.a}: the potential strength parameter in
eqn~\eqref{eqn:conforce} was set to $\alpha=25$, with simulation units defined
so that $m=1$, $k_\text{B}T=1$, $r_\text{c}=1$. A system of $N=250$ particles
was simulated in cubic periodic boundaries.  Timesteps in the range $0.005 \leq
\Delta t \leq 0.06$ were used, with run lengths up to 1000 reduced time units.
For the pairwise Nos\'{e}-Hoover thermostat, inertia parameters in the range
$0.2 \leq Q_\xi/N \leq 8.0$ were studied. For the configurational
Nos\'{e}-Hoover thermostat, inertia parameters in the range $2000 \leq Q_\mu/N
\leq 80000$ were used.

These thermostats allow one to check the accuracy of the integration by
monitoring the conserved energy-function eqns~\eqref{eqn:engxi} and
\eqref{eqn:engmu}. 
\begin{figure}[htp]
\caption{\label{fig:1}%
  Rate of change of energy-like function as a function of timestep $\Delta t$,
  plotted on log-log scales. Circles: pairwise Nos\'{e}-Hoover thermostat with
  inertia parameter $Q_\xi/N=0.8$. Squares: configurational Nos\'{e}-Hoover
  thermostat with inertia parameter $Q_\mu/N=2\times10^4$. Diamonds: velocity
  Verlet algorithm, with no
  thermostatting. The lines correspond to $\Delta t^4$ power law behaviour.}%
\includegraphics[width=0.5\textwidth,clip]{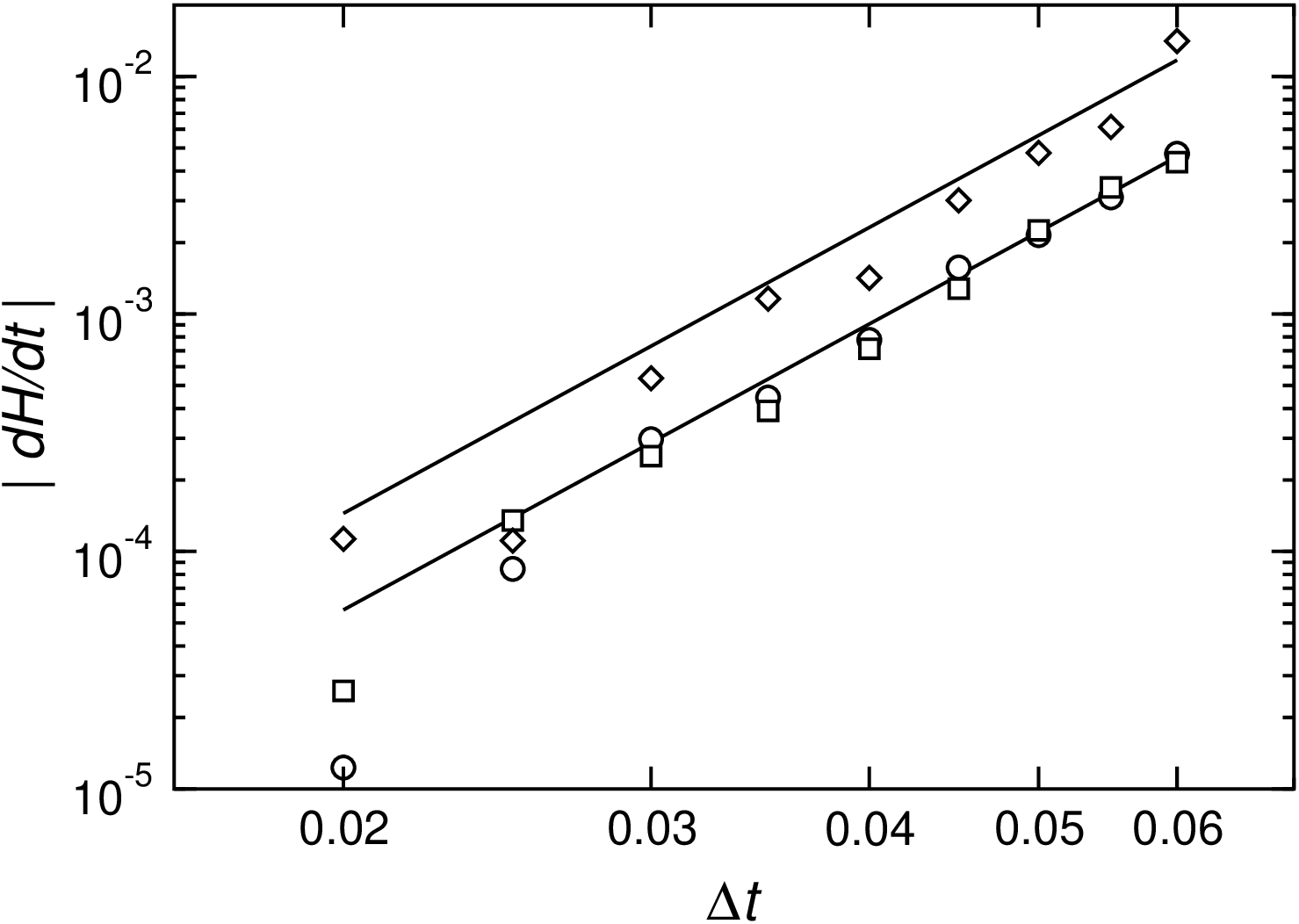}
\end{figure}
Figure \ref{fig:1} shows that, at timesteps $\Delta t > 0.02$ (very conservative
by DPD standards), there is a significant drift in this quantity: the rate of
increase is roughly proportional to $\Delta t^4$ at large $\Delta t$. This
problem has been noted before by \citet{hafskjold.b:2004.a}, and it is not
associated with the thermostatting, because the same behaviour is seen using the
simple velocity Verlet algorithm. The cause seems to be the relatively strong
discontinuity in force derivatives at the cutoff of the DPD potential
\citep{hafskjold.b:2004.a}. In DPD, and in MD with a thermostat, this tends to
be camouflaged. The present thermostats perform as well as (in fact, slightly
better than) velocity Verlet in this respect.

The oscillation of the internal energy of the particles (potential plus kinetic)
reflects the flow of energy into and out of the thermal reservoir, and this is
influenced by the choice of thermal inertia parameter. 
\begin{figure}[htp]
\caption{\label{fig:2}%
  Oscillation of internal energy $E=H$ as a function of time $t$.  Upper panel:
  pairwise Nos\'{e}-Hoover thermostat with inertia parameter: $Q_\xi/N=0.2$
  (solid line); $Q_\xi/N=0.8$ (dashed line); $Q_\xi/N=2.0$ (dot-dashed line.
  Lower panel: configurational Nos\'{e}-Hoover thermostat with inertia
  parameter: $Q_\mu/N=8\times10^3$ (solid line); $Q_\mu/N=2\times10^4$ (dashed
  line); $Q_\mu/N=4\times10^4$ (dot-dashed line).
}%
\includegraphics[width=0.5\textwidth,clip]{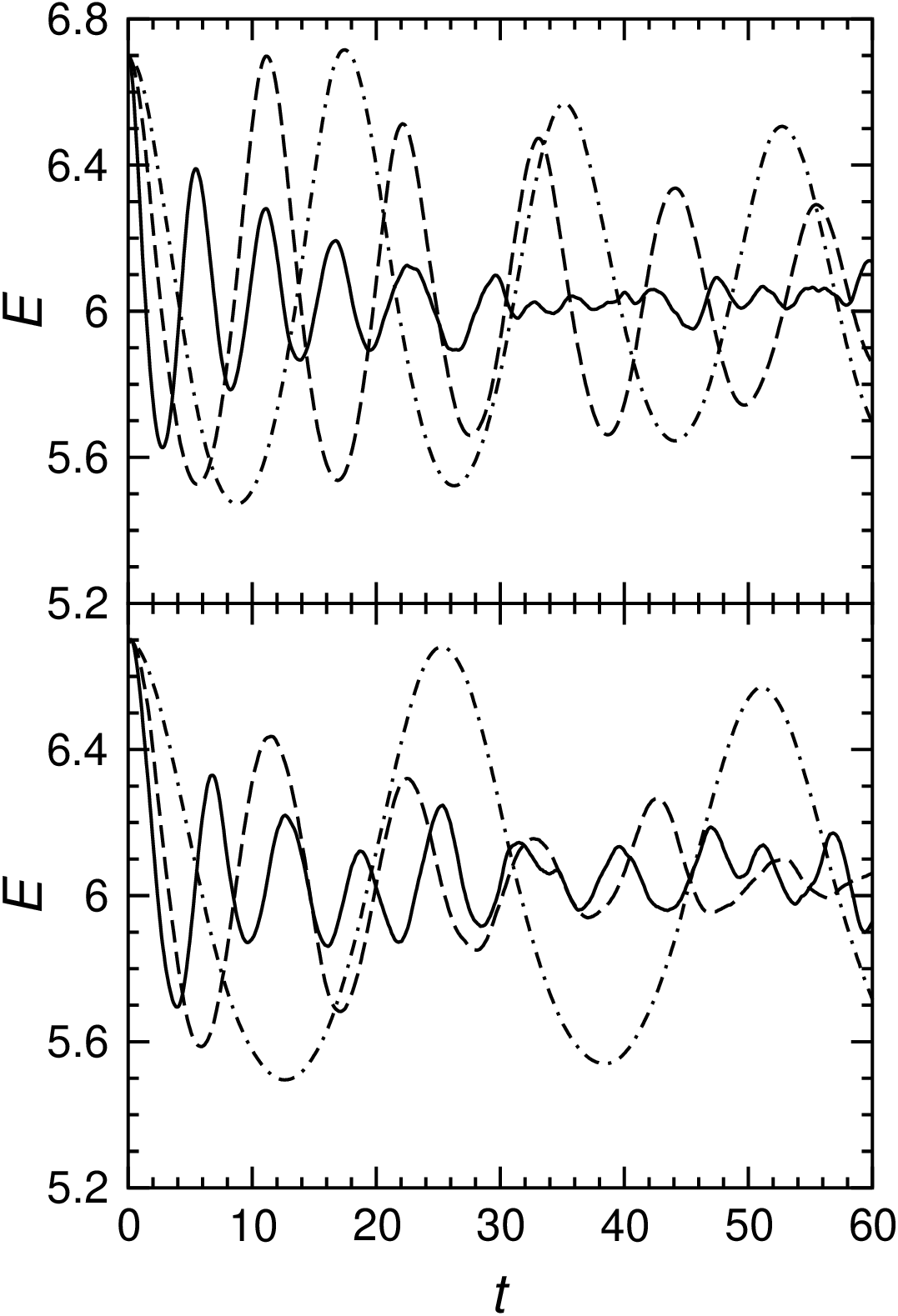}
\end{figure}
Typical results are shown in Fig.~\ref{fig:2}, and they show the expected
behaviour. There are damped oscillations: the runs lengths employed here are
typically long compared with the relaxation rate, while the timesteps are small
enough to cope with the oscillations. In this range, the precise choice of
thermal inertia is not critical.

\begin{figure}[htp]
\caption{\label{fig:3}%
  Kinetic temperature $T_\text{k}$ (open symbols) and configurational
  temperature $T_\text{c}$ (filled symbols) as functions of timestep $\Delta t$
  for three different thermostatting regimes:
(a)  pairwise Nos\'{e}-Hoover thermostat with inertia parameter
  $Q_\xi/N=0.4$;
(b) configurational Nos\'{e}-Hoover thermostat with inertia parameter
  $Q_\mu/N=4\times10^3$;
(c) both thermostats simultaneously.
}%
\includegraphics[width=0.6\textwidth,clip]{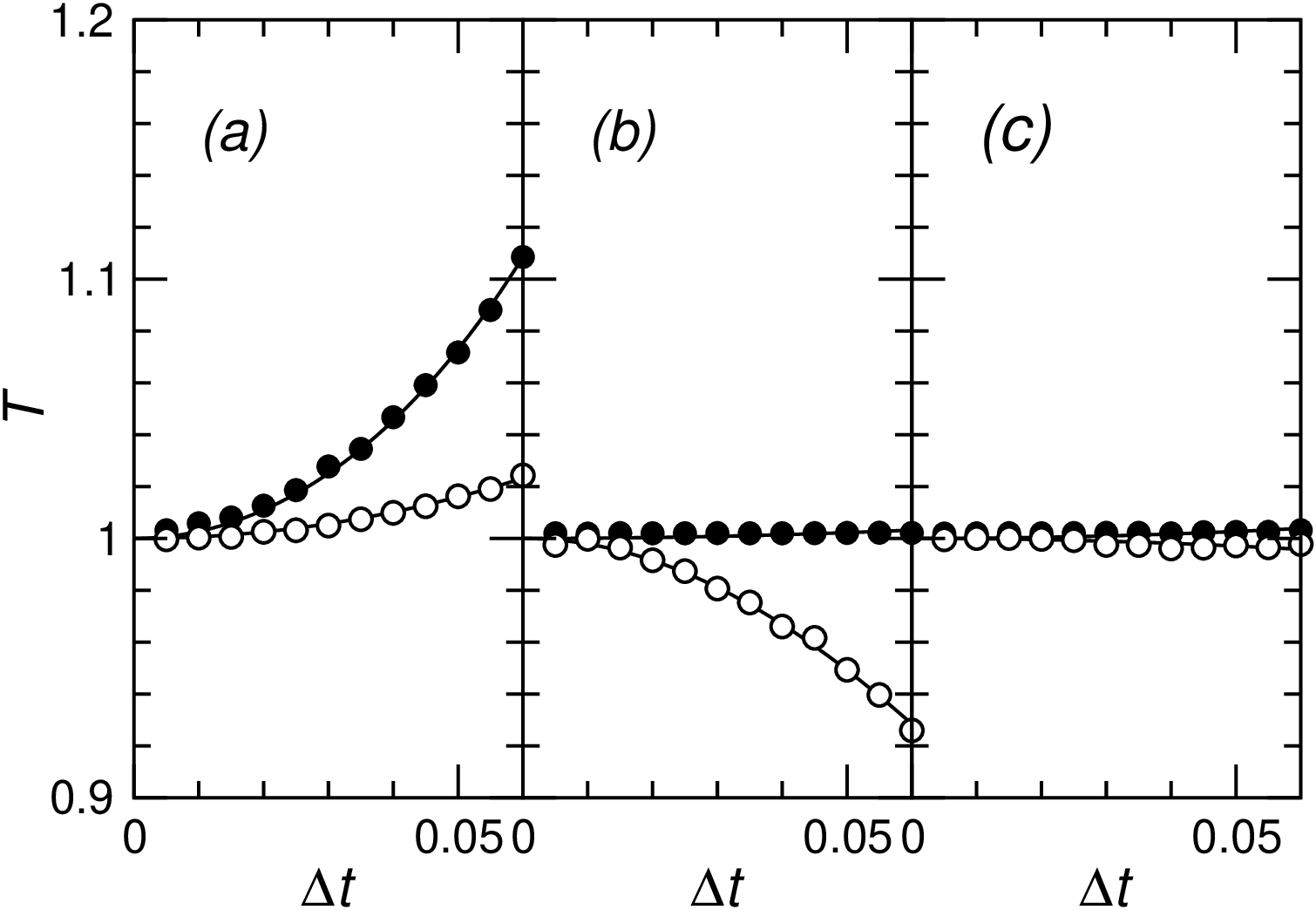}
\end{figure}

The simulation-averaged values of kinetic temperature $T_\text{k}$ (defined
through the total kinetic energy) and configurational temperature $T_\text{c}$
(defined by eqn~\eqref{eqn:conftemp}) are shown in Fig.~\ref{fig:3}.
When the kinetic temperature is controlled, lack of equilibrium is indicated by
the configurational temperature, which increases by as much as 10\% at the
largest timesteps studied. These results simply confirm what has been seen
before \citep{allen.mp:2006.a}: a measured kinetic temperature close to the
desired value should not be taken as a guarantee that the system is at
equilibrium. The form of the increase in $T_\text{c}$ may be understood
semi-quantitatively by considering the effect of non-zero-timestep
velocity-Verlet dynamics on the phase portrait of a simple harmonic oscillator
\citep{allen.mp:2006.a}. Conversely, when the configurational thermostat is
imposed, the measured kinetic temperature is significantly reduced when the
timestep is too large. This effect may be understood in a similar way by
considering harmonic oscillator velocity-Verlet dynamics: for a given positional
amplitude, the momentum amplitude is reduced as the timestep increases. When
both thermostats are applied together, not surprisingly, both $T_\text{c}$ and
$T_\text{k}$ are controlled well, up to the highest timesteps studied. This
deserves further investigation, but it would be over-optimistic to suppose that
the other degrees of freedom in the system are at equilibrium.

To illustrate the application to complex fluids, simulations of the same lipid
bilayer model studied previously
\cite{shillcock.jc:2002.a,jakobsen.af:2005.a,allen.mp:2006.a} have been carried
out with the new thermostat. Here, the solvent water is represented as before,
and each lipid molecule has the form of a 7-bead chain $\text{H}\text{T}_6$ in
which $\alpha$-repulsion parameters between hydrophilic ``head'' beads (H),
hydrophobic ``tail'' beads (T), and ``water'' beads (W) are chosen to produce
the desired behaviour \cite{shillcock.jc:2002.a}. Harmonic bond-stretching
potentials, and angle-bending potentials, act within the lipid molecules.
\begin{figure}[htp]
\caption{\label{fig:4}%
  Kinetic temperature $T_\text{k}$ (open symbols) and configurational
  temperature $T_\text{c}$ (filled symbols) as functions of timestep $\Delta t$
  for membrane simulations using three different thermostatting regimes:
(a)  pairwise Nos\'{e}-Hoover thermostat with inertia parameter
  $Q_\xi/N=0.4$;
(b) configurational Nos\'{e}-Hoover thermostat with inertia parameter
  $Q_\mu/N=4\times10^3$;
(c) both thermostats simultaneously.
Different symbols represent different DPD particle types:
circles, H, $\text{T}_6$;
squares, $\text{T}_1$, $\text{T}_5$;
diamonds, $\text{T}_2$, $\text{T}_3$, $\text{T}_4$;
triangles, water.
}%
\includegraphics[width=0.6\textwidth,clip]{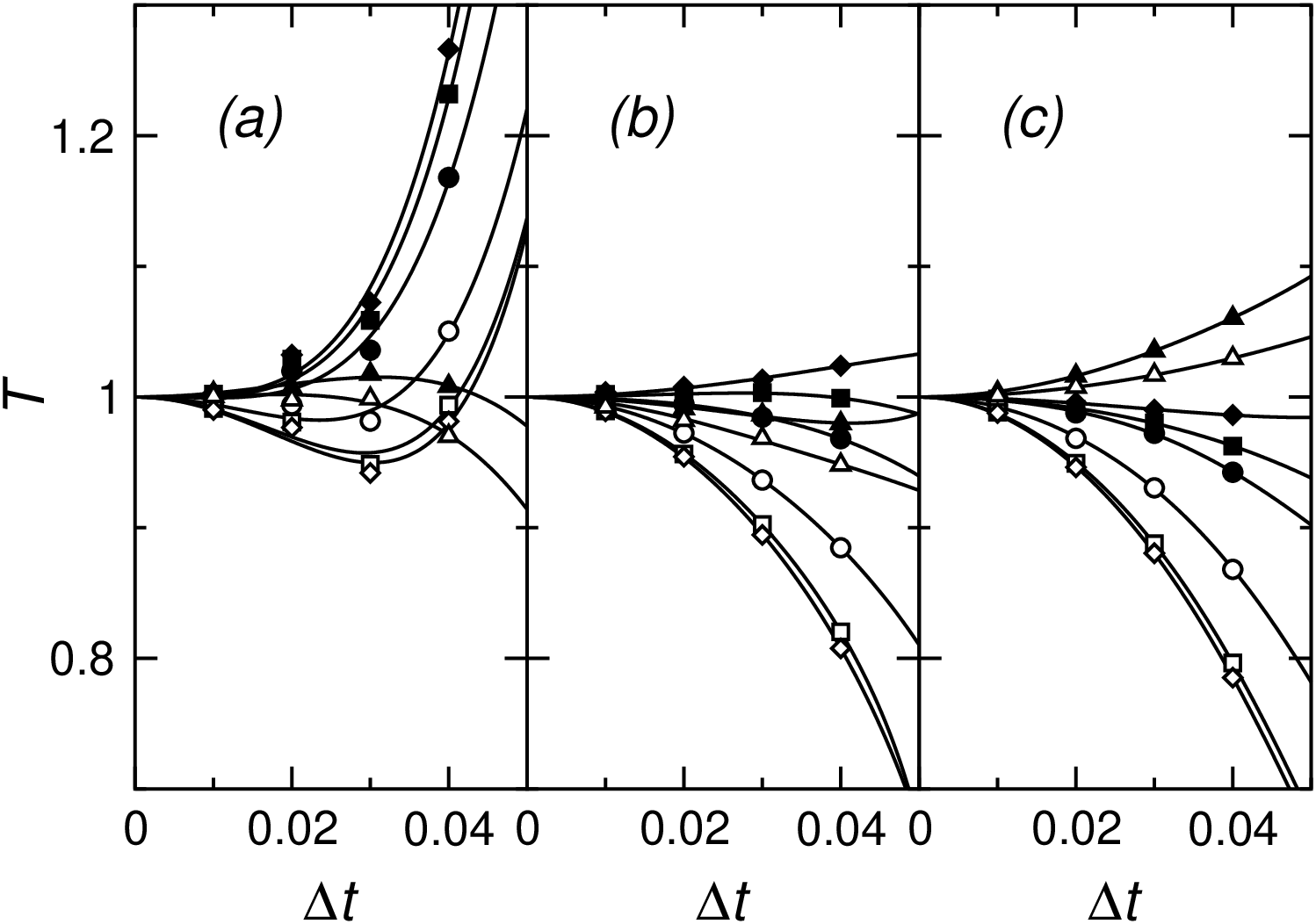}
\end{figure}
The measured temperatures of the different types of DPD bead are shown in
Figure~\ref{fig:4}. The results are consistent with those obtained before
\cite{allen.mp:2006.a} and show how dangerous it is to rely on thermostats to
equilibrate the system when the timestep is too large: the different bead types
have significantly different kinetic and configurational temperatures in all
cases. Actually, for this simple model, the cause of the problem, and the
remedy, are well understood. The \emph{intramolecular} potentials within the
lipid chains are too strong to be handled by the longer timesteps; this problem
is easily addressed by using multiple timestep methods
\citep{jakobsen.af:2006.a}. However, this example serves to illustrate possible
pitfalls which may occur in the general case.
\section{Discussion}
\label{sec:discussion}
The derivation of
Section~\ref{sec:thermostat} establishes the canonical ensemble as a stationary
distribution for the coordinates and momenta subject to the equations of motion
\eqref{eqn:nh}, although it does not prove that it is unique, nor guarantee that
a system will converge towards this distribution
\cite{hoover.wg:1991.a,kusnezov.d:1990.a,martyna.gj:1992.a}.  The result is
easily generalized to apply to a subset of pair interactions, simply by setting
$w_{ij}=0$ for the omitted pairs, making this suitable to combine with the Lowe
method as envisaged by \citet{stoyanov.sd:2005.a}. (Interestingly,
Ref.~\cite{hoover.wg:1991.a} contains exercises on incorporating a weighting
factor, and on considering a subset of degrees of freedom, for the conventional
Nos\'{e}-Hoover thermostat). Nos\'{e}-Hoover chains may easily be
added to further control the dynamics \citep{martyna.gj:1992.a}.

The preliminary results presented above indicate that the pairwise
Nos\'{e}-Hoover thermostat behaves as should be expected, and may be useful in
both DPD and conventional MD / NEMD simulations.  A feature of the proposed
thermostat, shared by the configurational-temperature thermostat, is the absence
of peculiar velocities: this may provide a more satisfactory way of controlling
the temperature than the conventional Nos\'{e}-Hoover thermostat in the case of
fluid flows, since only local relative velocities are used to define an
instantaneous temperature.
%
%
\begin{acknowledgments}
  The comments of Brad Holian, Bill Hoover, and Karl Travis on an early version
  of the manuscript are gratefully acknowledged. This work was conducted while
  MPA was on Study Leave at the University of Bielefeld. The research has been
  supported by the Engineering and Physical Sciences Research Council, and by
  the Alexander von Humboldt foundation.
\end{acknowledgments}
\bibliography{journals,main}

\providecommand{\acronym}[1]{\textsc{\MakeLowercase{#1}}}
\begin{thebibliography}{32}
\expandafter\ifx\csname natexlab\endcsname\relax\def\natexlab#1{#1}\fi
\expandafter\ifx\csname bibnamefont\endcsname\relax
  \def\bibnamefont#1{#1}\fi
\expandafter\ifx\csname bibfnamefont\endcsname\relax
  \def\bibfnamefont#1{#1}\fi
\expandafter\ifx\csname citenamefont\endcsname\relax
  \def\citenamefont#1{#1}\fi
\expandafter\ifx\csname url\endcsname\relax
  \def\url#1{\texttt{#1}}\fi
\expandafter\ifx\csname urlprefix\endcsname\relax\def\urlprefix{URL }\fi
\providecommand{\bibinfo}[2]{#2}
\providecommand{\eprint}[2][]{\url{#2}}

\bibitem[{\citenamefont{Nos\'e}(1984{\natexlab{a}})}]{nose.s:1984.a}
\bibinfo{author}{\bibfnamefont{S.}~\bibnamefont{Nos\'e}},
  \bibinfo{journal}{Molec.\ Phys.} \textbf{\bibinfo{volume}{52}},
  \bibinfo{pages}{255} (\bibinfo{year}{1984}{\natexlab{a}}).

\bibitem[{\citenamefont{Nos\'e}(1984{\natexlab{b}})}]{nose.s:1984.b}
\bibinfo{author}{\bibfnamefont{S.}~\bibnamefont{Nos\'e}}, \bibinfo{journal}{J.
  Chem.\ Phys.} \textbf{\bibinfo{volume}{81}}, \bibinfo{pages}{511}
  (\bibinfo{year}{1984}{\natexlab{b}}).

\bibitem[{\citenamefont{Andersen}(1980)}]{andersen.hc:1980.a}
\bibinfo{author}{\bibfnamefont{H.~C.} \bibnamefont{Andersen}},
  \bibinfo{journal}{J. Chem.\ Phys.} \textbf{\bibinfo{volume}{72}},
  \bibinfo{pages}{2384} (\bibinfo{year}{1980}).

\bibitem[{\citenamefont{Hoover}(1985)}]{hoover.wg:1985.a}
\bibinfo{author}{\bibfnamefont{W.~G.} \bibnamefont{Hoover}},
  \bibinfo{journal}{Phys.\ Rev.\ A} \textbf{\bibinfo{volume}{31}},
  \bibinfo{pages}{1695} (\bibinfo{year}{1985}).

\bibitem[{\citenamefont{Hoover}(1991)}]{hoover.wg:1991.a}
\bibinfo{author}{\bibfnamefont{W.~G.} \bibnamefont{Hoover}},
  \emph{\bibinfo{title}{Computational Statistical Mechanics}},
  vol.~\bibinfo{volume}{11} of \emph{\bibinfo{series}{Studies in Modern
  Thermodynamics}} (\bibinfo{publisher}{Elsevier}, \bibinfo{year}{1991}), ISBN
  \bibinfo{isbn}{0-444-88192-1}, \bibinfo{note}{available online at
  \url{http://williamhoover.info/book.pdf}}.

\bibitem[{\citenamefont{Delhommelle et~al.}(2003)\citenamefont{Delhommelle,
  Petravic, and Evans}}]{delhommelle.j:2003.a}
\bibinfo{author}{\bibfnamefont{J.}~\bibnamefont{Delhommelle}},
  \bibinfo{author}{\bibfnamefont{J.}~\bibnamefont{Petravic}}, \bibnamefont{and}
  \bibinfo{author}{\bibfnamefont{D.~J.} \bibnamefont{Evans}},
  \bibinfo{journal}{J. Chem.\ Phys.} \textbf{\bibinfo{volume}{119}},
  \bibinfo{pages}{11005} (\bibinfo{year}{2003}).

\bibitem[{\citenamefont{Evans and Morriss}(1986)}]{evans.dj:1986.a}
\bibinfo{author}{\bibfnamefont{D.~J.} \bibnamefont{Evans}} \bibnamefont{and}
  \bibinfo{author}{\bibfnamefont{G.~P.} \bibnamefont{Morriss}},
  \bibinfo{journal}{Phys.\ Rev.\ Lett.} \textbf{\bibinfo{volume}{56}},
  \bibinfo{pages}{2172} (\bibinfo{year}{1986}).

\bibitem[{\citenamefont{Evans et~al.}(1992)\citenamefont{Evans, T.Cui, Hanley,
  and Straty}}]{evans.dj:1992.a}
\bibinfo{author}{\bibfnamefont{D.~J.} \bibnamefont{Evans}},
  \bibinfo{author}{\bibfnamefont{S.}~\bibnamefont{T.Cui}},
  \bibinfo{author}{\bibfnamefont{H.~J.~M.} \bibnamefont{Hanley}},
  \bibnamefont{and} \bibinfo{author}{\bibfnamefont{G.~C.}
  \bibnamefont{Straty}}, \bibinfo{journal}{Phys.\ Rev.\ A}
  \textbf{\bibinfo{volume}{46}}, \bibinfo{pages}{6731} (\bibinfo{year}{1992}).

\bibitem[{\citenamefont{Travis et~al.}(1995)\citenamefont{Travis, Daivis, and
  Evans}}]{travis.kp:1995.b}
\bibinfo{author}{\bibfnamefont{K.~P.} \bibnamefont{Travis}},
  \bibinfo{author}{\bibfnamefont{P.~J.} \bibnamefont{Daivis}},
  \bibnamefont{and} \bibinfo{author}{\bibfnamefont{D.~J.} \bibnamefont{Evans}},
  \bibinfo{journal}{J. Chem.\ Phys.} \textbf{\bibinfo{volume}{103}},
  \bibinfo{pages}{10638} (\bibinfo{year}{1995}).

\bibitem[{\citenamefont{Stoyanov and Groot}(2005)}]{stoyanov.sd:2005.a}
\bibinfo{author}{\bibfnamefont{S.~D.} \bibnamefont{Stoyanov}} \bibnamefont{and}
  \bibinfo{author}{\bibfnamefont{R.~D.} \bibnamefont{Groot}},
  \bibinfo{journal}{J. Chem.\ Phys.} \textbf{\bibinfo{volume}{122}},
  \bibinfo{pages}{114112} (\bibinfo{year}{2005}).

\bibitem[{\citenamefont{Braga and Travis}(2005)}]{braga.c:2005.a}
\bibinfo{author}{\bibfnamefont{C.}~\bibnamefont{Braga}} \bibnamefont{and}
  \bibinfo{author}{\bibfnamefont{K.~P.} \bibnamefont{Travis}},
  \bibinfo{journal}{J. Chem.\ Phys.} \textbf{\bibinfo{volume}{123}},
  \bibinfo{pages}{134101/1} (\bibinfo{year}{2005}).

\bibitem[{\citenamefont{Hoogerbrugge and
  Koelman}(1992)}]{hoogerbrugge.pj:1992.a}
\bibinfo{author}{\bibfnamefont{P.~J.} \bibnamefont{Hoogerbrugge}}
  \bibnamefont{and} \bibinfo{author}{\bibfnamefont{J.~M. V.~A.}
  \bibnamefont{Koelman}}, \bibinfo{journal}{Europhys.\ Lett.}
  \textbf{\bibinfo{volume}{19}}, \bibinfo{pages}{155} (\bibinfo{year}{1992}).

\bibitem[{\citenamefont{Koelman and Hoogerbrugge}(1993)}]{koelman.jmva:1993.a}
\bibinfo{author}{\bibfnamefont{J.~M. V.~A.} \bibnamefont{Koelman}}
  \bibnamefont{and} \bibinfo{author}{\bibfnamefont{P.~J.}
  \bibnamefont{Hoogerbrugge}}, \bibinfo{journal}{Europhys.\ Lett.}
  \textbf{\bibinfo{volume}{21}}, \bibinfo{pages}{363} (\bibinfo{year}{1993}).

\bibitem[{\citenamefont{Espanol and Warren}(1995)}]{espanol.p:1995.b}
\bibinfo{author}{\bibfnamefont{P.}~\bibnamefont{Espanol}} \bibnamefont{and}
  \bibinfo{author}{\bibfnamefont{P.}~\bibnamefont{Warren}},
  \bibinfo{journal}{Europhys.\ Lett.} \textbf{\bibinfo{volume}{30}},
  \bibinfo{pages}{191} (\bibinfo{year}{1995}).

\bibitem[{\citenamefont{Lowe}(1999)}]{lowe.cp:1999.a}
\bibinfo{author}{\bibfnamefont{C.~P.} \bibnamefont{Lowe}},
  \bibinfo{journal}{Europhys.\ Lett.} \textbf{\bibinfo{volume}{47}},
  \bibinfo{pages}{145} (\bibinfo{year}{1999}).

\bibitem[{\citenamefont{Besold and Mouritsen}(2000)}]{besold.g:2000.a}
\bibinfo{author}{\bibfnamefont{G.}~\bibnamefont{Besold}} \bibnamefont{and}
  \bibinfo{author}{\bibfnamefont{O.~G.} \bibnamefont{Mouritsen}},
  \bibinfo{journal}{Comput.\ Mater.\ Sci.} \textbf{\bibinfo{volume}{18}},
  \bibinfo{pages}{225} (\bibinfo{year}{2000}).

\bibitem[{\citenamefont{Kusnezov et~al.}(1990)\citenamefont{Kusnezov, Bulgac,
  and Bauer}}]{kusnezov.d:1990.a}
\bibinfo{author}{\bibfnamefont{D.}~\bibnamefont{Kusnezov}},
  \bibinfo{author}{\bibfnamefont{A.}~\bibnamefont{Bulgac}}, \bibnamefont{and}
  \bibinfo{author}{\bibfnamefont{W.}~\bibnamefont{Bauer}},
  \bibinfo{journal}{Ann.\ Phys.} \textbf{\bibinfo{volume}{204}},
  \bibinfo{pages}{155} (\bibinfo{year}{1990}).

\bibitem[{\citenamefont{Martyna et~al.}(1992)\citenamefont{Martyna, Klein, and
  Tuckerman}}]{martyna.gj:1992.a}
\bibinfo{author}{\bibfnamefont{G.}~\bibnamefont{Martyna}},
  \bibinfo{author}{\bibfnamefont{M.~L.} \bibnamefont{Klein}}, \bibnamefont{and}
  \bibinfo{author}{\bibfnamefont{M.}~\bibnamefont{Tuckerman}},
  \bibinfo{journal}{J. Chem.\ Phys.} \textbf{\bibinfo{volume}{97}},
  \bibinfo{pages}{2635} (\bibinfo{year}{1992}).

\bibitem[{\citenamefont{Martyna et~al.}(1994)\citenamefont{Martyna, Tobias, and
  Klein}}]{martyna.gj:1994.a}
\bibinfo{author}{\bibfnamefont{G.~J.} \bibnamefont{Martyna}},
  \bibinfo{author}{\bibfnamefont{D.~J.} \bibnamefont{Tobias}},
  \bibnamefont{and} \bibinfo{author}{\bibfnamefont{M.~L.} \bibnamefont{Klein}},
  \bibinfo{journal}{J. Chem.\ Phys.} \textbf{\bibinfo{volume}{101}},
  \bibinfo{pages}{4177} (\bibinfo{year}{1994}).

\bibitem[{\citenamefont{Martyna et~al.}(1996)\citenamefont{Martyna, Tuckerman,
  Tobias, and Klein}}]{martyna.gj:1996.a}
\bibinfo{author}{\bibfnamefont{G.~J.} \bibnamefont{Martyna}},
  \bibinfo{author}{\bibfnamefont{M.~E.} \bibnamefont{Tuckerman}},
  \bibinfo{author}{\bibfnamefont{D.~J.} \bibnamefont{Tobias}},
  \bibnamefont{and} \bibinfo{author}{\bibfnamefont{M.~L.} \bibnamefont{Klein}},
  \bibinfo{journal}{Molec.\ Phys.} \textbf{\bibinfo{volume}{87}},
  \bibinfo{pages}{1117} (\bibinfo{year}{1996}).

\bibitem[{\citenamefont{Hirschfelder}(1960)}]{hirschfelder.jo:1960.a}
\bibinfo{author}{\bibfnamefont{J.~O.} \bibnamefont{Hirschfelder}},
  \bibinfo{journal}{J. Chem.\ Phys.} \textbf{\bibinfo{volume}{33}},
  \bibinfo{pages}{1462} (\bibinfo{year}{1960}).

\bibitem[{\citenamefont{Rugh}(1997)}]{rugh.hh:1997.a}
\bibinfo{author}{\bibfnamefont{H.~H.} \bibnamefont{Rugh}},
  \bibinfo{journal}{Phys.\ Rev.\ Lett.} \textbf{\bibinfo{volume}{78}},
  \bibinfo{pages}{772} (\bibinfo{year}{1997}).

\bibitem[{\citenamefont{Butler et~al.}(1998)\citenamefont{Butler, Ayton, Jepps,
  and Evans}}]{butler.bd:1998.a}
\bibinfo{author}{\bibfnamefont{B.~D.} \bibnamefont{Butler}},
  \bibinfo{author}{\bibfnamefont{G.}~\bibnamefont{Ayton}},
  \bibinfo{author}{\bibfnamefont{O.~G.} \bibnamefont{Jepps}}, \bibnamefont{and}
  \bibinfo{author}{\bibfnamefont{D.~J.} \bibnamefont{Evans}},
  \bibinfo{journal}{J. Chem.\ Phys.} \textbf{\bibinfo{volume}{109}},
  \bibinfo{pages}{6519} (\bibinfo{year}{1998}).

\bibitem[{\citenamefont{Han and Grier}(2004)}]{han.yl:2004.a}
\bibinfo{author}{\bibfnamefont{Y.~L.} \bibnamefont{Han}} \bibnamefont{and}
  \bibinfo{author}{\bibfnamefont{D.~G.} \bibnamefont{Grier}},
  \bibinfo{journal}{Phys.\ Rev.\ Lett.} \textbf{\bibinfo{volume}{92}},
  \bibinfo{pages}{148301} (\bibinfo{year}{2004}).

\bibitem[{\citenamefont{Han and Grier}(2005)}]{han.yl:2005.a}
\bibinfo{author}{\bibfnamefont{Y.~L.} \bibnamefont{Han}} \bibnamefont{and}
  \bibinfo{author}{\bibfnamefont{D.~G.} \bibnamefont{Grier}},
  \bibinfo{journal}{J. Chem.\ Phys.} \textbf{\bibinfo{volume}{122}},
  \bibinfo{pages}{064907} (\bibinfo{year}{2005}).

\bibitem[{\citenamefont{Allen}(2006)}]{allen.mp:2006.a}
\bibinfo{author}{\bibfnamefont{M.~P.} \bibnamefont{Allen}},
  \bibinfo{journal}{J. Phys.\ Chem.\ B} \textbf{\bibinfo{volume}{110}},
  \bibinfo{pages}{3823} (\bibinfo{year}{2006}).

\bibitem[{\citenamefont{Delhommelle and Evans}(2001)}]{delhommelle.j:2001.a}
\bibinfo{author}{\bibfnamefont{J.}~\bibnamefont{Delhommelle}} \bibnamefont{and}
  \bibinfo{author}{\bibfnamefont{D.~J.} \bibnamefont{Evans}},
  \bibinfo{journal}{Molec.\ Phys.} \textbf{\bibinfo{volume}{99}},
  \bibinfo{pages}{1825} (\bibinfo{year}{2001}).

\bibitem[{\citenamefont{Groot and Warren}(1997)}]{groot.rd:1997.a}
\bibinfo{author}{\bibfnamefont{R.~D.} \bibnamefont{Groot}} \bibnamefont{and}
  \bibinfo{author}{\bibfnamefont{P.~B.} \bibnamefont{Warren}},
  \bibinfo{journal}{J. Chem.\ Phys.} \textbf{\bibinfo{volume}{107}},
  \bibinfo{pages}{4423} (\bibinfo{year}{1997}).

\bibitem[{\citenamefont{Hafskjold et~al.}(2004)\citenamefont{Hafskjold, Liew,
  and Shinoda}}]{hafskjold.b:2004.a}
\bibinfo{author}{\bibfnamefont{B.}~\bibnamefont{Hafskjold}},
  \bibinfo{author}{\bibfnamefont{C.~C.} \bibnamefont{Liew}}, \bibnamefont{and}
  \bibinfo{author}{\bibfnamefont{W.}~\bibnamefont{Shinoda}},
  \bibinfo{journal}{Molec.\ Simul.} \textbf{\bibinfo{volume}{30}},
  \bibinfo{pages}{879} (\bibinfo{year}{2004}).

\bibitem[{\citenamefont{Shillcock and Lipowsky}(2002)}]{shillcock.jc:2002.a}
\bibinfo{author}{\bibfnamefont{J.~C.} \bibnamefont{Shillcock}}
  \bibnamefont{and} \bibinfo{author}{\bibfnamefont{R.}~\bibnamefont{Lipowsky}},
  \bibinfo{journal}{J. Chem.\ Phys.} \textbf{\bibinfo{volume}{117}},
  \bibinfo{pages}{5048} (\bibinfo{year}{2002}).

\bibitem[{\citenamefont{Jakobsen et~al.}(2005)\citenamefont{Jakobsen,
  Mouritsen, and Besold}}]{jakobsen.af:2005.a}
\bibinfo{author}{\bibfnamefont{A.~F.} \bibnamefont{Jakobsen}},
  \bibinfo{author}{\bibfnamefont{O.~G.} \bibnamefont{Mouritsen}},
  \bibnamefont{and} \bibinfo{author}{\bibfnamefont{G.}~\bibnamefont{Besold}},
  \bibinfo{journal}{J. Chem.\ Phys.} \textbf{\bibinfo{volume}{122}},
  \bibinfo{pages}{204901} (\bibinfo{year}{2005}).

\bibitem[{\citenamefont{Jakobsen et~al.}(2006)\citenamefont{Jakobsen, Besold,
  and Mouritsen}}]{jakobsen.af:2006.a}
\bibinfo{author}{\bibfnamefont{A.~F.} \bibnamefont{Jakobsen}},
  \bibinfo{author}{\bibfnamefont{G.}~\bibnamefont{Besold}}, \bibnamefont{and}
  \bibinfo{author}{\bibfnamefont{O.~G.} \bibnamefont{Mouritsen}},
  \bibinfo{journal}{J. Chem.\ Phys.} \textbf{\bibinfo{volume}{124}},
  \bibinfo{pages}{094104} (\bibinfo{year}{2006}).

\end{thebibliography}
\end{document}